\documentstyle[12pt]{article}

\def\beq{\begin{equation}}
\def\eeq{\end{equation}}
\def\bea{\begin{eqnarray}}
\def\eea{\end{eqnarray}}
\def\nn{\nonumber}
\begin{document}
\begin{titlepage}
\begin{center}
{{\it P.N.Lebedev Institute preprint} \hfill FIAN/TD-9/91\\
{\it I.E.Tamm Theory Department} \hfill ITEP-M-8/91
\begin{flushright}{October 1991}\end{flushright}
\vspace{0.1in}{\Large\bf  Unification of All String Models with $c<1$}\\[.4in]
{\large  S.Kharchev, A.Marshakov, A.Mironov}\\
\bigskip {\it P.N.Lebedev Physics
Institute \\ Leninsky prospect, 53, Moscow, 117 924}
\footnote{E-mail address: theordep@sci.fian.msk.su},\\ \smallskip
\bigskip {\large A.Morozov}\\
 \bigskip {\it Institute of Theoretical and Experimental
Physics,  \\
 Bol.Cheremushkinskaya st., 25, Moscow, 117 259}\\ \smallskip
\bigskip {\large A.Zabrodin}\\
 \bigskip {\it Institute of Chemical Physics\\ Kosygina st., 117334, Moscow}}
\end{center}
\bigskip
\bigskip

\centerline{\bf ABSTRACT}
\begin{quotation}

A 1-matrix model is proposed, which nicely interpolates between double-scaling
continuum limits of all multimatrix models. The interpolating partition
function is always a KP $\tau $-function and always obeys
${\cal L}_{-1}$-constraint and string equation.
Therefore this model can be considered as a natural
unification of all models of 2d-gravity (string models) with $c\leq 1.$
\end{quotation}
\end{titlepage}

\setcounter{page}1
{\it The model.} The purpose of this letter is to introduce a new theory, which
we call {\it Generalized Kontsevich's Model} (GKM) and to describe its
structure and appealing properties. The partition function of the GKM is
defined by the following integral over $N \times N$ Hermitean matrix:

\beq
Z^{\{{\cal V}\}}_N[M] \equiv  \frac{\int e^{U(M,Y)}dY}
{\int e^{-U_2(M,Y)}dY} \hbox{ , }
\eeq
where

\beq
U(M,Y) = Tr[{\cal V}(M+Y) - {\cal V}(M) - {\cal V}\ '(M)Y]
\eeq
and

\beq
U_2(M,Y) =   \lim_{\epsilon \rightarrow 0}
{1\over \epsilon ^2} U(M,\epsilon Y) \hbox{ , }
\eeq
is an $Y^2$-term in $U$. $M$ is also a Hermitean $N \times N$ matrix with
eigenvalues $\{\mu _i\}$,
${\cal V}(\mu)$ is arbitrary analytic function.

\bigskip
{\it Integrable structure.} After the shift of variables $X=Y+M$ and
integration over angular components of $X$,
$Z^{\{{\cal V}\}}_N[M]$ acquires the form of

\beq
Z^{\{{\cal V}\}}_N[M] = {[\det
\tilde \Phi _i(\mu_j)]\over \Delta (M)}\hbox{,}
\eeq
where  $\Delta (M) = \prod _{i<j}(\mu_i-\mu_j)$  is the Van-der-Monde
determinant, and functions

\beq
\tilde \Phi _i(\mu) = [{\cal V}\ ''(\mu)]^{1/2}
e^{{\cal V}(\mu)-\mu{\cal V}\ '(\mu)}\int
e^{-{\cal V}(x)+x{\cal V}'(\mu)} x^i dx
\eeq
The only assumption necessary for the derivation of (4) from (1) is the
possibility to represent the potential  ${\cal V}(\mu)$  as a formal series
in positive {\it integer} powers of $\mu$.

Formula (4) with {\it arbitrary} entries  $\phi _i(\mu)$  is characteristic
for generic KP $\tau $-function  $\tau ^G(T_n)$  in Miwa's coordinates

\beq
T_n = {1\over n} TrM^{-n}\hbox{ , } n ge 1
\eeq
and the point $G$ of Grassmannian is defined by potential $\cal V$
through the set of basis vectors
$\{\phi _i(\mu)\}$. (We remind that {\it a priori} definition
is  $\tau ^G(T_n) = <0| e^{\sum \ T_nJ_n} G|0>,$ where  $J$  stands for the
free-fermion $U(1)$ current and $G$ is an exponent of quadratic combination of
free fermion operators.) Therefore

\beq
Z^{\{{\cal V}\}}[M] = \tau ^{\{{\cal V}\}}(T_n).
\eeq
The case of finite $N$ in this formalism is distinguished by the condition that
only $N$ of the parameters $\{\mu_i\}$ are finite. In order to take the
limit $N \rightarrow  \infty $  in the GKM (1) it is enough to bring all the
$\mu_i's$ from infinity. In this sense this a smooth limit, in contrast
to the singular conventional double-scaling limit, which one needs to take in
ordinary (multi)matrix models.

\bigskip
${\cal L}_{-1}$-{\it constraint.} The set of function
$\{\tilde \Phi _i({\mu})\}$ in (4) is, however, not arbitrary. They are all
expressed through a single function --- potential ${\cal V}({\mu}),- and$ are
in fact recurrently related: if we denote the integral in (5) through
$F_i({\cal V}\ '({\mu}))$, then

\beq
F_i(\lambda ) = (\partial /\partial \lambda )^{i-1}F_1(\lambda )\hbox{.}
\eeq
This relation is enough to prove, that

\beq
{\partial \over \partial T_1}\log \ Z^{\{{\cal V}\}}_N
 = - Tr\ M + Tr {\partial \over \partial \Lambda _{tr}}
\log \ \det \ F_i(\lambda _j)
\eeq
whenever potential ${\cal V}({\mu})$ grows faster than $\mu$ as $\mu
\rightarrow  \infty .$

Thus, $Z ^{\{{\cal V}\}}$ satisfies a simple identity:

\beq
{1\over Z^{\{{\cal V}\}}} {\cal L}^{\{{\cal V}\}}_{-1}Z^{\{{\cal V}\}}_N =
{\partial \over \partial T_1}\log \ Z^{\{{\cal V}\}}_N + TrM -
Tr{\partial \over \partial \Lambda _{tr}}\log \ \det \ F_i(\lambda _j) = 0
\eeq
where operator ${\cal L}^{\{{\cal V}\}}_{-1}$ is defined to be

\bea
{\cal L}^{\{{\cal V}\}}_{-1}= \sum _{n\geq 1}Tr
[{1\over {\cal V}\ ''(M)M^{n+1}}] {\partial \over \partial T_n} +
\nn \\
+ {1\over 2} \sum _{i,j}{1\over {\cal V}\ ''({\mu}_i){\cal V}\ ''({\mu}_j)}
{{\cal V}\ ''({\mu}_i)-{\cal V}\ ''({\mu}_j)\over {\mu}_i -  {\mu}_j} +
{\partial \over \partial T_1}
\eea
(the items with $i=j$ are included into the sum). The reason why this operator
is denoted by ${\cal L}_{-1}$ will be clear after
reductions of GKM will be discussed. From eqs.(9),(10) it follows, that
partition function of GKM usually satisfies the constraint

\beq
{\cal L}^{\{{\cal V}\}}_{-1}\tau ^{\{{\cal V}\}} = 0.
\eeq

\bigskip
{\it Reductions.} The integral  ${\cal F}^{\{{\cal V}\}}[\Lambda ]$,  $\Lambda
\equiv  {\cal V}\ '(M)$, in the numerator of (1) satisfies the Ward identity

\beq
Tr\left\lbrace \epsilon (\Lambda )\left[ {\cal V}\ '({\partial \over %
\partial \Lambda _{tr}}) - \Lambda \right] \right\rbrace
{\cal F}^{\{{\cal V}\}}_N =  0
\eeq
(as result of invariance under any shift of integration variable $X
\rightarrow  X + \epsilon (M))$. If potential ${\cal V}({\mu})$ is restricted
to be a polynomial of degree $K+1$, this identity implies, that the functions
(8) obey additional relations:

\beq
F_{m+Kn}(\lambda) =
\lambda ^n\cdot F_m(\lambda) +
\sum ^{m+Kn-1}_{i=1}s_iF_i(\lambda) \hbox{ . }
\eeq
Since the sum at the $r.h.s$. does not contribute to determinant (5), we can
say
that all the functions  $F_n$ are expressed through the first  $K$  functions
$F_1...F_K$ by multiplication by powers of  $\lambda = {\cal V}\ '({\mu})$.
Such
situation (when the basis vectors  $\phi _i$, defining the point of
Grassmannian are proportional to the first  $K$ ones) corresponds to reduction
of KP-hierarchy. This reduction depends on the form of  ${\cal V}\ '({\mu})$
and in the case of ${\cal V}({\mu})  = {\cal V}_K({\mu}) =
const\cdot {\mu}^{K+1}$ coincides with the well-known  $K$-reduction of the
KP-hierarchy (KdV as  $K=2$, Boussinesq as  $K=3$ etc.). Thus in such cases
partition function of GKM becomes  $\tau ^{\{K\}}$-function of the
corresponding hierarchy. Generic $\tau ^{\{K\}}$ possesses an important
property: it is
almost independent of all time-variables  $T_{nK}$. To be exact,

\beq
\partial \ log\tau ^{\{K\}}/\partial T_{nK} = a_n = const
\eeq
If ${\cal V} = {\cal V}_K$, the generic expression (12) for the
${\cal L}_{-1}$-operator turns into

\beq
{\cal L}^{\{K\}}_{-1} = {1\over K} \sum _{n>K} nT_n\partial /\partial T_{n-K} +
{1\over 2K }\sum_{{a+b=K}\atop {a,b>0}}
aT_abT_b + \partial /\partial T_1
\eeq
The last item at the $r.h.s$. may be eliminated by the shift of time-variables:

\beq
T_n \rightarrow  \hat T^{\{K\}}_n = T_n + {K\over n} \delta _{n,K+1}.
\eeq
This shift is, however, $K$-dependent and does not seem to have too much sense.
However, only expressed in terms of  these $\hat T$'s the constraint (12)
acquires the
form of

\bea
{\cal L}^{\{K\}}_{-1}\tau ^{\{K\}} = \left\lbrace {1\over K} \sum _{{n>K} \atop
{n \ne 0 mod K}}
n\hat T_n\partial /\partial \hat T_{n-K} + {1\over 2K}
\sum _{{a+b=K}\atop {a,b>0}}a\hat T_ab\hat T_b\right\rbrace  \tau ^{\{K\}} =\nn
\\
= \sum  _n a_n(n+1)\hat T_{(n+1)K} \tau ^{\{K\}} \hbox{ . }
\eea
with the $l.h.s.$ familiar from [1].
The sum at the $r.h.s$. of (18) does not contribute to the ``string equation"

\beq
{\partial \over \partial T_1}
{{\cal L}^{\{K\}}_{-1}\tau ^{\{K\}}\over \tau ^{\{K\}}} = 0 \hbox{ . }
\eeq
Moreover, in variance with generic $\tau ^{\{K\}}$ the partition function
$Z ^{\{K\}}$ of GKM
is expected to obey (15) and (18) with all $a_n=0$.

\bigskip
{\it Universal string equation.} Generalization of (19) to the case of
arbitrary potential

\beq
{\partial \over \partial T_1}
{{\cal L}^{\{\cal V\}}_{-1}\tau ^{\{\cal V\}}\over \tau ^{\{\cal V\}}} = 0
\hbox{ . }
\eeq
may be transformed to the following form

\beq
\sum_{n\ge -1} {\cal T}_n \frac{\partial ^2 \log \tau}{\partial T_1
\partial T_n} = u \hbox{ , }
\eeq
where

\bea
{\cal T}_n \equiv Tr {1 \over V''(M)}{1 \over M^{n+1}} \hbox{ , } \\
u \equiv \frac{\partial ^2 \log \tau }{(\partial T_1 )^2}\hbox{ , }
\frac{\partial \log \tau }{\partial T_0} \equiv 0 \hbox{ , }
\frac{\partial \log \tau }{\partial T_{-1}} \equiv T_1 \hbox{ . } \nn
\eea
If Baker-Akhiezer are introduced:

\beq
\Psi _{\pm} (z|T_k) = e^{\sum T_k z^k}\frac{\tau (T_n \pm {z^{-n} \over n})}
{\tau (T_n)} \hbox{ , }
\eeq
string equation (22) can be rewritten in the form of bilinear relation

\beq
\sum_i \frac{\Psi _+ (\mu_i) \Psi _-(\mu _i)}{\mu _i} = u \hbox{ . }
\eeq

\bigskip
$\cal W${\it -constraints.} According to the arguments of refs.[1] the
constraint

\beq
{\cal L}^{\{K\}}_{-1}\tau ^{\{K\}} = 0
\eeq
$(i.e$. (18) with the vanishing $r.h.s.$, as it is in fact the case if we deal
with the model (1)) implies the entire tower of $\cal W$-constraints

\beq
{\cal W}^{(k)}_{Kn} Z^{\{K\}} = 0\hbox{, }   k = 2,3,...,K\hbox{; }  n\geq  1-k
\eeq
imposed on  $\tau ^{\{K\}}$. Here ${\cal W}^{(p)}_{Kn}$ is the $n-th$ harmonics
of the  $p-th$ generator of Zamolodchikov's  $W_K$-algebra (the proper notation
would be  ${\cal W}^{(p)\{K\}}_n$, but it is a bit too complicated). There is a
Virasoro Lie sub-algebra, generated by ${\cal W}^{(2)}_{Kn} =
{\cal L}^{\{K\}}_n$, and the particular ${\cal L}^{\{K\}}_{-1}$ is just the
operator (16). This is the origin of our notation
${\cal L}^{\{{\cal V}\}}_{-1}$ in the generic situation (where the entire
Virasoro subalgebra of $W_\infty $ was not explicitly specified).

Besides being a corollary of (24), the constraints (25) can be directly deduced
from the Ward identity (13). For the case of  $K=2$  (which is original
Kontsevich's model [2]) this derivation was given in ref.[3] (see also [4,5]
for
alternative proofs). Unfortunately, for $K\geq 3$ the direct corollary of
(13) is not just (25), but peculiar linear combinations of these constraints,
$e.g$. for  $K=3$  they look like

\bea
{\cal W}^{(3)}_{3n}Z^{\{3\}}_\infty  = 0\hbox{, }     n\geq -2;
\nn \\
\left\lbrace \sum _{k\geq 1}(3k-1)\hat T_{3k-1}{\cal W}^{(2)}_{3k+3n}
+\sum _{a+b=3n} {\partial \over \partial T_{3a+2}}
{\cal W}^{(2)}_{3b-3}\right\rbrace  Z^{\{3\}}_\infty  = 0\hbox{, }
a,b\geq 0\hbox{, }  n\geq -2;
\nn \\
\left\lbrace \sum _{k\geq 1+\delta _{n+3,0}}(3k-2)\hat T_{3k-2}%
{\cal W}^{(2)}_{3k+3n} +\sum _{a+b=3n}
{\partial \over \partial T_{3a+1}} {\cal W}^{(2)}_{3b-3}\right\rbrace
Z^{\{3\}}_\infty  = 0\hbox{, }  a,b\geq 0\hbox{, }  n\geq -3.
\eea
For identification of (26) with (25) one can argue, that both sets of
constraints possess unique, and thus coinciding, solutions.

\bigskip
{\it Multimatrix models.} While detailed investigation of the properties of
multimatrix models in the double-scaling limit (the analogue of ref.[6] in the
case of conventional Hermitean 1-matrix model) is still lacking, it has been
suggested in [1] that the square roots of their partition functions,
$\sqrt{\Gamma ^{\{K-1\}}_{ds}}$ ( $K-1$ is the number of matrices, index
{\it ds} means, that partition function is considered in the double scaling
limit), possess the following properties:

\beq
{\cal W}^{(k)}_{Kn}\sqrt{\Gamma ^{\{K-1\}}_{ds}} = 0\hbox{, }   k =
2,3,...,K\hbox{; }   n \geq  1-k.
\eeq
Comparing these properties to the above information about GKM, we obtain:

\beq
Z^{\{K\}}_\infty  = \sqrt{\Gamma ^{\{K-1\}}_{ds}}
\eeq

\bigskip
{\it Conclusion.} To conclude, we presented a brief description of the
properties of the GKM, defined by eq.(1). Its partition function may be
considered as a functional of two different variables:  potential
${\cal V}({\mu})$ and the infinite-dimensional Hermitean matrix $M$ with
eigenvalues $\{{\mu}_i\}$. Partition function $Z^{\{{\cal V}\}}_N$ is an
$N$-independent KP $\tau $-function, considered as a function of time-variables
$T_n= {1\over n}TrM^{-n}$ and the point of Grassmannian is specified by the
choice of potential. The $N$-dependence enters only through the argument  $M:$
we return to finite-dimensional matrices if only $N$ eigenvalues of $M$ are
finite. In this sense the ``continuum" limit of  $N \rightarrow  \infty $  is
smooth.

The GKM is associated with a subset of Grassmannian, specified by additional
${\cal L}_{-1}$-constraint (12). For particularly adjusted potentials
${\cal V}({\mu}) = const\cdot {\mu}^{K+1}$, the corresponding points in
Grassmannian lies in the subvarieties, associated with $K$-reductions of
KP-hierarchy,  $Z^{\{{\cal V}\}}$ becomes independent of all the time-variables
$T_{Kn}$, and the  ${\cal L}_{-1}$-constraint implies the whole tower of
$W_K$-algebra constraints on the reduced $\tau $-function. These properties are
exactly the same as suggested for double scaling limit of the $K-1$-matrix
model, and in fact there is an identification (29).

All this means, that GKM provides an interpolation between double-scaling
continuum limits of all multimatrix models and thus between all string models
with $c\leq 1$. Moreover, this is a reasonable interpolation, because both
integrable and ``string-equation" structures are preserved. This is why we
advertise GKM as a plausible (on-shell) prototype of a unified theory of 2d
gravity. All the proofs will be presented in ref.[7].

\bigskip
We are grateful to A.Gerasimov, M.Kontsevich and Yu.Makeenko for
useful discussions.
\bigskip
\begin{center}{{\Large\bf References}}\end{center}
\medskip
1. M.Fukuma, H.Kawai, R.Nakayama  Int.J.Mod.Phys. A6 (1991) 1385\\
2. M.Kontsevich  Funk.Anal. i Priloz. 25 (1991) 50\\
3. A.Marshakov, A.Mironov, A.Morozov preprint HU-TFT-91-44,
ITEP-M-4/91, \\ FIAN/TD/04-91\\
4. Yu.Makeenko, G.Semenoff  ITEP/UBC preprint, July 1991\\
5. E.Witten  in talk at NYC conference, June 1991\\
6. Yu.Makeenko et.al. Nucl.Phys. B356(1991) 574\\
7. S.Kharchev et al. Preprint ITEP-M-9/91 --- FIAN/TD-10/91
\end{document}